\documentclass[preprint,12pt]{elsarticle}

\usepackage{color}
\usepackage{amssymb}
\usepackage{amsmath}
\usepackage{url}
\usepackage{graphicx}
\usepackage[lined,boxed,linesnumbered]{algorithm2e}

\journal{}

\begin{document}

\begin{frontmatter}



\title{Heterogeneous information network model for equipment-standard system}


\author[add1]{Liang Yin}
\author[add1]{Li-Chen Shi}
\author[add1]{Jun-Yan Zhao}
\author[add1]{Song-Yang Du}
\author[add3,add4]{Wen-Bo Xie}
\author[add2,add3,add4]{Duan-Bing Chen\corref{cor1}}
\ead{dbchen@uestc.edu.cn}

\address[add1]{Beijing Special Vehicle Institute, Beijing 100072, People's Republic of China}
\address[add2]{The Center for Digitized Culture and Media, UESTC, Chengdu 611731, People's Republic of China}
\address[add3] {Big Data Research Center, University of Electronic Science and Technology of China, Chengdu 611731, People's Republic of China}
\address[add4]{Web Sciences Center, University of Electronic Science and Technology of China, Chengdu 611731, People's Republic of China.}

\cortext[cor1]{Corresponding author at: Web Sciences Center, University of Electronic Science and Technology of China, Chengdu 611731, People's Republic of China.}
\begin{abstract}
Entity information network is used to describe structural relationships between entities. Taking advantage of its extension and heterogeneity, entity information network is more and more widely applied to relationship modeling. Recent years, lots of researches about entity information network modeling have been proposed, while seldom of them concentrate on equipment-standard system with properties of multi-layer, multi-dimension and multi-scale. In order to efficiently deal with some complex issues in equipment-standard system such as standard revising, standard controlling, and production designing, a heterogeneous information network model for equipment-standard system is proposed in this paper. Three types of entities and six types of relationships are considered in the proposed model. Correspondingly, several different similarity-measuring methods are used in the modeling process. The experiments show that the heterogeneous information network model established in this paper can reflect relationships between entities accurately. Meanwhile, the modeling process has a good performance on time consumption.
\end{abstract}

\begin{keyword}
Complex system \sep Heterogeneous information network \sep Equipment-standard system \sep  Entity relationships model


\end{keyword}

\end{frontmatter}


\section{Introduction}
\label{intro}
Complex network theory has been proven to be a powerful framework to understand the structure and dynamics of complex systems\cite{ref1,ref2,ref3,ref4,ref5,ref6,ref7}. Entity information network is a kind of complex network that describes the structural relationships between entities. With more and more researches about entity information network model having been proposed, it is widely used in social network analyzing\cite{ref8,ref9,ref10,ref11}, image association and other fields \cite{ref12,ref13}. Depending on the diversity of the relationships and entities, entity information network model can be divided into two categories: based on simple structure and based on complex one.

Generally, only one kind of entity relationship is contained in simple structure based model, such as item-to-item, object-to-item or object-to-object relationships.

In recent years, many modeling methods on item-to-item information network are proposed. In refs \cite{ref14,ref15}, the similarity between media sources is evaluated by mapping different types of medium's features to a common space based on media contextual clues. Zhu et al. \cite{ref16} proposed an information network model to correlate tweets, emotion features and users based on emotion analysis.

Matrix transformation, matrix decomposition and random walk methods are used to construct object-to-item information network \cite{ref17,ref18, ref19}. And some researches focus on the status of users in the resource, such as influence, importance and opinion leaders, by constructing the authoritative network based on specific topics \cite{ref20}.
Some modeling methods are also widely concentrated on object-to-object relationships such as user-to-user. Based on direct and indirect users' relationships, the similarity between users can be measured by the comments on social media resource and those on user reviews \cite{ref21,ref22}. Besides, matrix decomposition can also be used to evaluate the similarity between users in social media combining with LDA \cite{ref23}.

Complex structure based model supports multiple forms of relationships between multimodal entities. Google's Knowledge Graph \cite{ref24} and other engine knowledge maps belong to this type of model. In which, there are multi-class of relationships between entities. In addition, the entities are also multi-dimension and multi-scale. Thus, they are generally called heterogeneous information network models \cite{ref25}.

Although heterogeneous information network models are widely studied in complex systems, such as academic resource search \cite{ref26}, citations recommendation \cite{ref27}, user based personalized service \cite{ref28,ref29}, traveling plan search and recommendation \cite{ref30}, and makeup recommendation \cite{ref31}. There are seldom researches about equipment-standard system.

As we know, equipment-standard system is a multi-layer, multi-dimension and multi-scale complex system. For this reason, we present a heterogeneous information network model for equipment-standard system (HINM-ESS) in this paper. HINM-ESS contains three types of nodes that present different granularity of entities in equipment-standard system and six types of entity relationships. A complete HINM-ESS can provide strong support for equipment-standard system, such as resource searching, production designing, standard revising and controlling.

Two real data sets are used in experiments to verify the validity of HINM-ESS. The one is a real equipment-standard system data set that contains 2600 standard documents and 24 elements. The other is a mixed test data set that contains different size of data from multiple fields. The experiments show that our methods in modeling process are efficient and accurate. Comparing with Word Mover's Distance (WMD) \cite{ref34}, the relational modeling between documents using our method can save 50\% time and the performance of precision reduces about 20\%. That is, we can establish HINM-ESS efficiently, and reflect relationship between entities in equipment-standard system accurately.

\section{ Model and Method}
\subsection{Framework of HINM-ESS}
The formal expression of HINM-ESS is described as HINM-ESS=$(V,E)$. In which, $V=\{Doc,Item,Topic\}$ is the network node set with three different granularity of entities, i.e., $Doc$, $Item$ and $Topic$. $Doc$ represents the standard document; $Item$ represents the clause in standard document; $Topic$ represents the unit such as equipment, module or element. $E=\{E_{DD},E_{DI},E_{DT},E_{II},E_{IT},E_{TT} \}$ is the network edge set with six different kinds of relationships between entities, where $E_{DD}$ represents the $Doc-Doc$ relationships, $E_{DI}$ represents the $Doc-Item$ relationships, $E_{DT}$ represents the $Doc-Topic$ relationships, $E_{II}$ represents the $Item-Item$ relationships, $E_{IT}$ represents the $Item-Topic$ relationships, and $E_{TT}$ represents the $Topic-Topic$ relationships. Each edge has its weight to measure the degree of correlation or the similarity between entities.

\begin{figure}
  \centering
  \includegraphics[width=14cm]{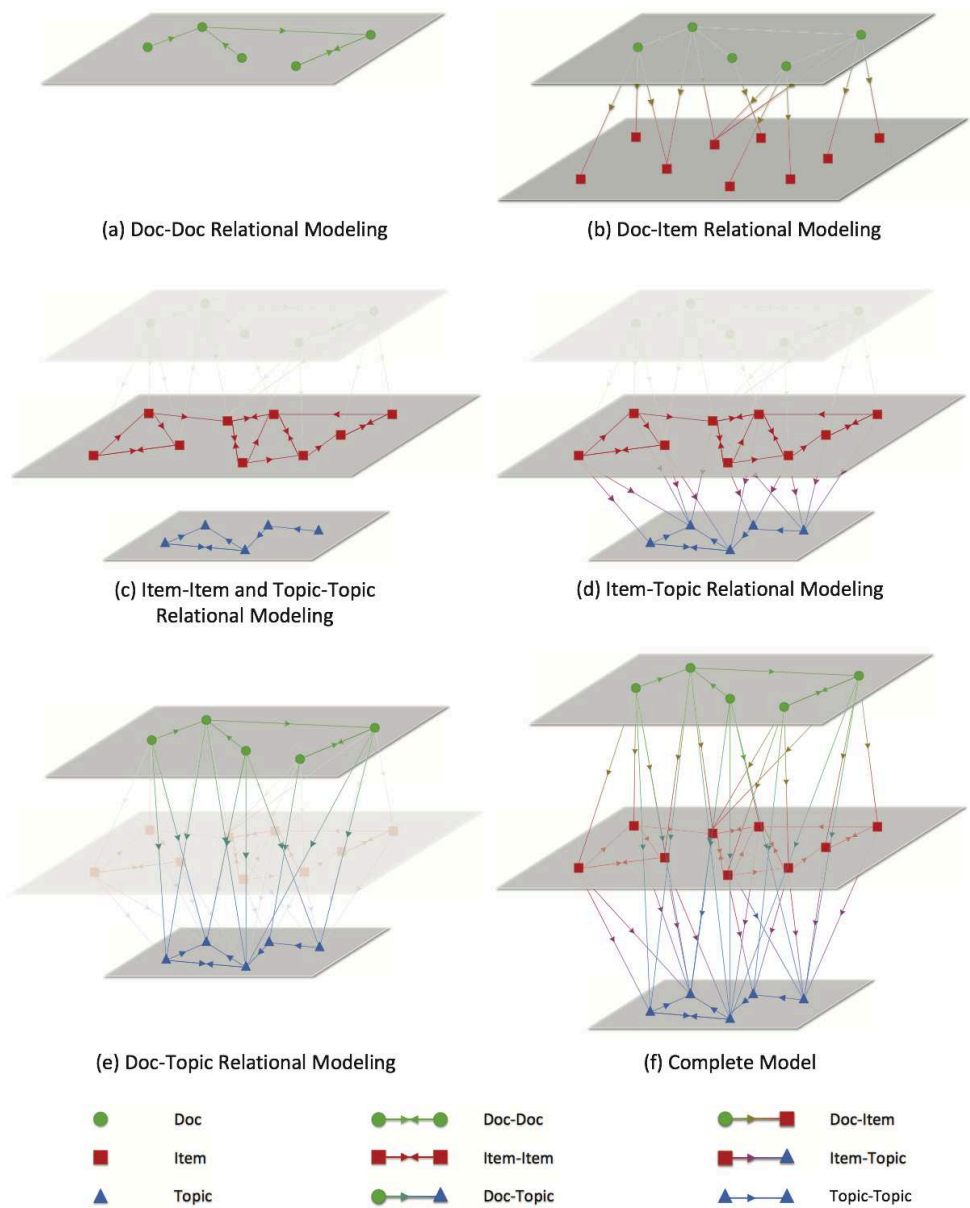}\\
  \caption{The framework of HINM-ESS modeling.}\label{Fig1}
\end{figure}

As shown in Fig. \ref{Fig1}, to construct the HINM-ESS, six kinds of entity relationships are confirmed in turn by evaluating the similarities or correlations between entities. First of all, weights of $E_{DD}$ are evaluated to confirm the $Doc-Doc$ relationships, as shown in Fig. \ref{Fig1}(a). Secondly, a $Doc$ is divided into several items to confirm $Doc-Item$ relationships $E_{DI}$, as shown in Fig. \ref{Fig1}(b). And then $Item-Item$ relationships $E_{II}$ and $Topic-Topic$ relationships $E_{TT}$ are confirmed by the same strategy used in the first step, as shown in Fig. \ref{Fig1}(c). Thirdly, the $Item-Topic$ relationships $E_{IT}$ can be confirmed since $Item-Item$ and $Topic-Topic$ relationships have been confirmed, as shown in Fig. \ref{Fig1}(d), and $Doc-Topic$ relationships $E_{DT}$ can be confirmed in virtue of $E_{DD}$ and $E_{TT}$ , as shown in Fig. \ref{Fig1}(e). Finally, HINM-ESS can be obtained, as shown in Fig. \ref{Fig1}(f). Because these six relationships involve different entities, the method to measure the weights of edges are very different. The details will be described in following subsections.

\subsection{ $Doc-Doc$ relational modeling}
The $Doc-Doc$ relational modeling is the first and the most important step to establish the HINM-ESS. As the contents of $Docs$ are text, the correlation between $Docs$ can be confirmed via text similarity so as to establish the connection between two $Doc$ nodes in HINM-ESS. Since the contents of $Items$ and $Topics$ are text as well, only $Doc-Doc$ relational modeling is described in details, the $Item-Item$ and $Topic-Topic$ relational modeling is similar with $Doc-Doc$.

\subsubsection{Docs similarity calculation via WMD}

Text analyzing is one of the most popular research topics. There are many researches focus on measuring text similarity, such as LDA \cite{ref32} and word2vec \cite{ref33}. These models translate text contents to different abstract features to improve the measuring performance. In this paper, we choose Word Mover's Distance (WMD) \cite{ref34} method to measure the similarity between $Docs$.

WMD provides accurate similarity measurement by combining the word embedding with the Earth Mover's Distance (EMD). In WMD, word2vec provides embedding matrix $\mathbf{X} \in \mathbb{R}^{d\times n}$ for a finite size vocabulary of $n$ words. The $i^{th}$ column, $\overrightarrow{x}_i\in \mathbf{R}^d$ , represents the embedding of the $i^{th}$ word in $d-$ dimensional space. Therefore, semantic similarity between two words $w_i$ and $w_j$, that is refer to as the cost associated with `traveling' from one word to another, is measured by their Euclidean distance in the word2vec embedding space, that is,
\begin{equation}\label{eq1}
  cost_{ij}=\|\overrightarrow{x}_i-\overrightarrow{x}_j\|_2.
\end{equation}

Text documents are represented as normalized bag-of-words (nBOW) vectors, $\overrightarrow{d}\in \mathbf{R}^n$, if word $w_i$ appears $c_i$ times in the document, the $i^{th}$ element ${d}_i$ in $\overrightarrow{d}$ is defined as,
\begin{equation}\label{eq2}
 d_i=\frac{c_i}{\sum_{j=1}^n c_j}.
\end{equation}

Let $\overrightarrow{d}$ and $\overrightarrow{d'}$ be the nBOW representation of two text documents, $T\in \mathbb{R}^{n\times n}$ be a flow matrix denotes `how much' of word $w_i$ in $\overrightarrow{d}$ travels to word $w_j$ in $\overrightarrow{d'}$, The similarity evaluation problem can be defined as the minimum cumulative cost required to move all words from $\overrightarrow{d}$ to $\overrightarrow{d'}$, and the constraints are provided by the solution to the following linear program,
\begin{eqnarray}\label{eq3}
   & & \min\limits_{T_{ij}\geq 0}\sum_{i,j=1}^n{T_{ij}\cdot cost_{ij}},\\ \nonumber
  subject\quad to & & \sum_{j=1}^nT_{ij}=d_{i} \quad \forall i\in \{1,2,\cdots ,n\}, \\
   & & \sum_{i=1}^nT_{ij}=d'_{j}\quad \forall j\in \{1,2,\cdots ,n\}. \nonumber
\end{eqnarray}

After the similarities between $Docs$ being calculated by WMD, we regard these similarity values as the weights of $E_{DD}$.

Although WMD leads to low error rates, the time complexity is as high as $O(n^3\log n)$. In order to improve the time consumption while maintaining the accuracy, we modify WMD by introducing SimHash \cite{ref35} strategy. In which, Top$-N$ potentially similar $Docs$ are screened via SimHash to reduce the complexity significantly.

\subsubsection{ Top-N potentially similar Docs screening}
The main idea of SimHash strategy is to reduce the dimension. As a local sensitive Hash method, SimHash maps the high-dimensional eigenvectors to a signature value containing $f$ bits named $f-$bit fingerprint. By comparing the Hamming distance between $f-$bit fingerprints, we can estimate whether the $Docs$ are similar or not. Therefore, an efficient potentially similar $Docs$ screening strategy based on SimHash is presented in this paper.

In Top$-N$ potentially similar $Docs$ screening process, SimHash is used to map $f-$bit fingerprints for each $Doc$. The $i^{th}$ $Doc$ $D_i$ in dataset is mapped to $f-$bit fingerprints with number 0 or 1, for instance,
$\mathop {\overbrace{(0101...00111)}}\limits^{f\quad bits}$. Based on the fingerprints, the Hamming distance is chosen to estimate the similarity between $Docs$ and generate a list of Top$-N$ potentially similar $Docs$ efficiently. As shown in Eq. \ref{eq4}, Hamming distance is the number of `1' in XOR between documents $D_n$ and $D_m$.
\begin{equation}\label{eq4}
  H_{nm}  = D_n  \oplus D_m.
\end{equation}

For instance, the Hamming distance between 110 and 011 equals 2 since $110\oplus 011=101$.

After estimating the similarity between $Docs$ by Hamming distance, we can obtain the Top$-N$ potentially similar documents list by directly using sorting algorithms. But this simple strategy is very time-consuming. Even for the best sorting algorithm, its time complexity reaches $O(n \log n)$. Aiming at the Top$-N$ $Docs$, there is no need to sort all Hamming distances. Therefore, we propose two strategies to improve the efficiency of potentially similar document lists generating process.

\textbf{(1) The lowliest replace elimination based strategy.}

In this strategy, Top$-N$ potentially similar documents are stored in a finite set of $k$ elements, which is defined as $SET=\{D_i|0 \leq i<k\}$. The update strategy of $SET$ is

\begin{equation}\label{eq5}
SET=\left\{
\begin{aligned}
  \left( SET\bigcup D_\Delta  \right)\backslash \left\{ \max (SET) \right\} &,& Ham(\max (SET)) > Ham(D_\Delta  )  \\
  SET &,& Ham(\max (SET)) \le Ham(D_\Delta  )  \\
\end{aligned}
\right.
\end{equation}
where $\max(SET)$ represents the node which is the farthest from the target node on hamming distance in $SET$; $D_\Delta$ represents a new node which is under judgment to be the potentially similar $Docs$; $Ham(\cdot)$ represents the hamming distance between this node and target node.

\textbf{(2) Ordered window filling based strategy.}

The lowliest replace elimination based strategy needs to repeatedly update the maximum value in the finite set. It takes too much time to traverse the whole set. Therefore, we propose another strategy based on ordered window filling to reduce traversal time further. In this strategy, Top$-N$ potentially similar documents are stored in an ordered window of $k$ elements, which is defined as $WIN = \{ D_i|0\leq i < k, Ham(D_i) \leq Ham(D_{i + 1})\}$. So, the elements $D_i$ in $WIN$ can be updated via Eq. \ref{eq6} under the condition of $Ham(D_m )\leq Ham(D_\Delta )\leq Ham(D_{m+1})$,
\begin{equation}\label{eq6}
D_i = \left\{
\begin{aligned}
   D_{i-1}&,&m+1< i \leq k  \\
   D_\Delta &,& i=m+1  \\
   D_i&, &0\leq i<m  \\
\end{aligned} \right.
\end{equation}
in which, $D_\Delta$ represent a new node which is under judgment to be the potentially similar $Docs$; $k$ is the size of $WIN$.

\begin{figure}
  \centering
  \includegraphics[width=14cm]{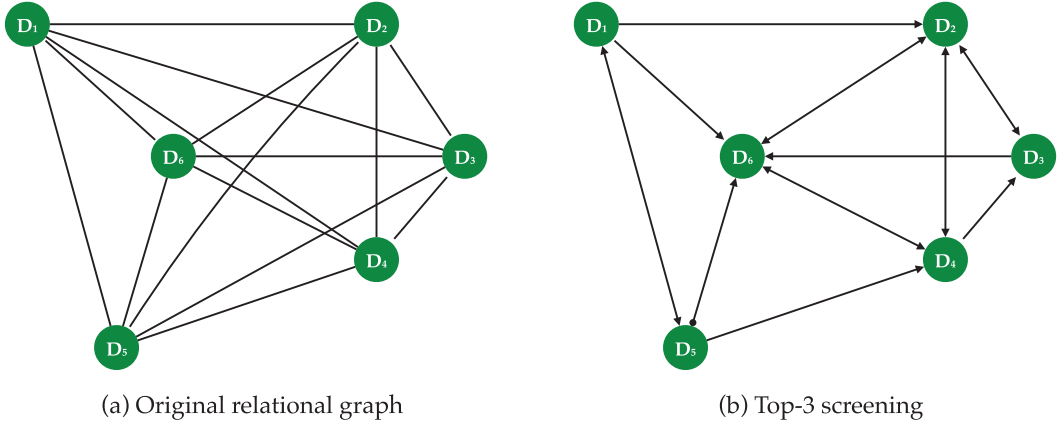}\\
  \caption{$Doc-Doc$ relational modeling processes.}\label{Fig2}
\end{figure}

Via the screen strategies above, we obtain the Top$-N$ potentially similar $Docs$ by removing weightless edges from original $Doc-Doc$ relational graph. This process is illustrated by the simple example in Fig. \ref{Fig2}.

Firstly, we have a lot of edges in the original $Doc-Doc$ relational graph that is a complete graph, as shown in Fig. \ref{Fig2}(a). By screening the Top-3 potentially similar $Docs$, we remove the weightless edges in Fig. \ref{Fig2}(a), and then obtain the $Doc-Doc$ relational graph, as shown in Fig. \ref{Fig2}(b). The edges are directed since two $Docs$ may not be the Top$-N$ similar for each other at the same time.

As there are only a small number of edges which link the potentially similar documents left, the time consumption of the similarity evaluation process with WMD method is reduced sharply. Then $Doc-Doc$ relational graph can be completed efficiently via two stages, i.e., Top$-N$ potentially similar $Doc$ screening and Docs' similarity evaluation via WMD method.

\subsection{$Doc-Item$ Relational modeling}
Each $Doc$ contains different numbers of $Items$. Therefore, the $Doc-Item$ relational modeling can be considered as a decomposition process. The core issue of the $Doc-Item$ relational modeling process is to extract items from standard documents accurately.

Since $Docs$ in equipment-standard system have typical hierarchical structure, the section numbers in section headers are always composed of integers and symbolic points, `2.2' for instance. Furthermore, the section number will be followed by the section title directly. Therefore, all the possible section numbers and titles can be extracted via regular expressions to construct a triplet sequence $T_i=(Cap_i, No_i, In_i)$. In which, $Cap$ is the chapter number that is the first integer of section number (for example, `2' is the chapter number when the section number is `2.2'); $No$ is the whole section number; $In$ is the line number of title in the document. For instance, a possible triplet may like this: (5, 5.2.1, 456).

The results obtained via regular expressions still contain some noises that meet the definition of the section title but not the real one, such as data in the table or in the text of the reference data. For this reason, we design following noise-filtering rule to remove those illogical triplets,

\begin{equation}\label{eq7}
\left\{
\begin{aligned}
   No_{i-1}  &  <  No_{i }   \\
   \max \left\{ {Cap_j |j < i} \right\} &  \leq  Cap_i ,  \\
   In_{i - 1}  &  <  In_i   \\
\end{aligned} \right.
\end{equation}
in which, section number must conform the typesetting format of sections  (when $No_{i-1}=2.2$, the logical value of $No_i$ can only be 2.2.1 or 2.3 or 3); the chapter number $Cap_i$ must be no less than all the chapter numbers in the previous triplets. $In_i$ in the triplet sequence must be sorted.

According to the noise-filtering rule, most of noises are erased. As a result, we can use the line number in triplets to decompose the $Docs$ and to extract out the $Items$ accurately. At the same time, the $Doc-Item$ relational network is constructed.

79 standard documents have been used to test the $Item$ extracting process. There are 77 documents extracted correct completely. The other two documents have only 1 incorrect item respectively. The accuracy of the $Items$ extracting process is 97.5\%.

\subsection{$Item-Topic$ relational modeling}
In HINM-ESS, two relationships, $Doc-Topic$ and $Item-Topic$, are widely used in equipment-standard system application. Just $Item-Topic$ relational modeling process is analyzed in this paper, since $Doc-Topic$ relational modeling is similar.

Generally, there are part of $Items$ have been assigned $Topic$ labels, that is, there are some known $Item-Topic$ relationships in equipment-standard system. Hence, these relationships can be used to measure the correlation between $Items$ and $Topics$ indirectly, and then complete the $Item-Topic$ relational model.

For a new $Item$ $I_i$ that is waiting for assigning $Topic$ label, the correlation $W_{I_i,T_k}$ between $I_i$ and the $k^{th}$ $Topic$ $T_k$ is measured by
\begin{equation}\label{eq8}
  W_{I_i,T_k}=\frac{\sum\limits_{I_j\in S_{I_i}} W_{I_i,I_j}\cdot W_{I_j,T_k}}{|S_{I_i}|},
\end{equation}
where $I_j$ is one of $I_i$'s similar $Items$; $W_{I_i,I_j}$ is the similarity between $I_i$ and its similar $Item$ $I_j$ which obtained in $Item-Item$ relational modeling; $S_{I_i}$ presents $I_i$'s similar $Items$ set; $W_{I_j,T_k}$ presents the correlations between $I_i$'s similar $Items$ and their relative $Topics$.

Analogously, for a new $Topic$ $T_k$, the correlation between $T_k$ and the $i^{th}$ $Item$ $I_i$ is measured by combining with $T_k$'s similar $Topics$ and their relative $Items$,
\begin{equation}\label{eq9}
W_{I_i,T_k}=\frac{\sum\limits_{T_p\in S_{T_k}} W_{I_i,T_p}\cdot W_{T_p,T_k}}{|S_{T_k}|},
\end{equation}
where $T_p$ is one of $T_k$'s similar $Topics$; $W_{T_p,T_k}$ is the similarity between $T_k$ and its similar $Topic$ $T_p$; $S_{T_k}$ presents $T_p$'s similar $Topics$ set; $W_{I_i,T_p}$ is the correlations between $T_k$'s similar $Topics$ and their relative $Items$.

\begin{figure}
  \centering
  \includegraphics[width=14cm]{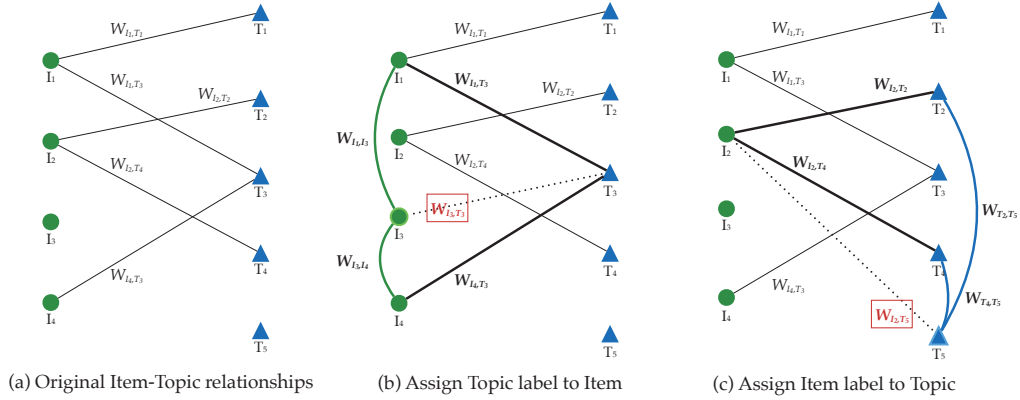}\\
  \caption{Examples of $Item-Topic$ relational modeling.}\label{Fig3}
\end{figure}

The $Item-Topic$ relational modeling process, which is the core of the HINM-ESS, is illustrated in Fig. \ref{Fig3}. Figure \ref{Fig3}(a) shows four $Items$, five $Topics$ and five known relationships. In original graph, $I_3$ and $T_5$ are waiting for assignment. Since $I_1$ and $I_4$ are assigned to $T_3$, the correlation between $I_3$ and $T_3$ is measured by

\begin{equation}
   W_{I_3,T_3}=\frac{W_{I_3,I_1}\cdot W_{I_1,T_3}+W_{I_3,I_4}\cdot W_{I_4,T_3}}{2},
\end{equation}
as shown in Fig. \ref{Fig3}(b). Similarly, the correlations $W_{I_2,T_2}$ and $W_{I_2,T_4}$ are known, therefore, the correlation between $I_2$ and $T_5$ is measured by,
\begin{equation}
  W_{I_2,T_5}=\frac{W_{I_2,T_2}\cdot W_{T_2,T_5}+W_{I_2,T_4}\cdot W_{T_4,T_5}}{2},
\end{equation}
as shown in Fig. \ref{Fig3}(c).

After all the correlations are evaluated, the weightless $Item-Topic$ edges will be removed according to a threshold $\theta$. Alike soft-classification method, multiple $Topics$ will be assigned to one $Item$ in $Item-Topic$ relational modeling process and vice versa.

In the $Item-Topic$ relational model, if the relationship between a $Item$ and $Topic$ is confirmed, they will be put into the training samples. This iterative process will optimize HINM-ESS continually and make $Items-Topic$ model more and more accurate.

\section{Results and Discussions}
In this paper, the performance of $Doc-Doc$ and $Item-Topic$ relational network modeling methods are tested on two real data (DATA1 and DATA2). DATA1 includes 2600 standard documents and 24 elements. DATA2 is a mixed-field text data set which contains different size of data (from 1KB to 1GB).

The performance of original WMD and SimHash+WMD is compared on DATA1 through three indices. The first index is Time Improvement ($TI$)
 \begin{equation}\label{eq10}
   TI=\frac{T_{WMD}-T_{SimHash+WMD}}{T_{WMD}},
 \end{equation}
where $T_{WMD}$ represents the time consumption of WMD algorithm and $T_{SimHash+WMD}$ represents the time consumption of SimHash+WMD method.

The second index is accuracy that is used to estimate the proportion of documents SimHash mistakenly delete from the 20 most similar text evaluated by WMD,

\begin{equation}\label{eq11}
   Accuracy_N=\frac{|\min\limits_{20}(WMD)\bigcap \min\limits_{20}(SimHash+WMD)|}{20},
 \end{equation}
where $Accuracy_N$ is the accuracy when SimHash method screening top$-N$ documents as potentially similar documents; $\min\limits_{20}(WMD)$ is the set that contains the 20 most similar documents evaluated by WMD and $\min\limits_{20}(SimHash+WMD)$ is that by SimHash+WMD.

The third index is $F_1-score$ that considers the time improvement and accuracy at the same time. It is defined as
\begin{equation}\label{eq12}
   F_1-score=\frac{2\times TI \times Accuracy}{TI + Accuracy}.
 \end{equation}

\begin{figure}
  \centering
  \includegraphics[width=10cm]{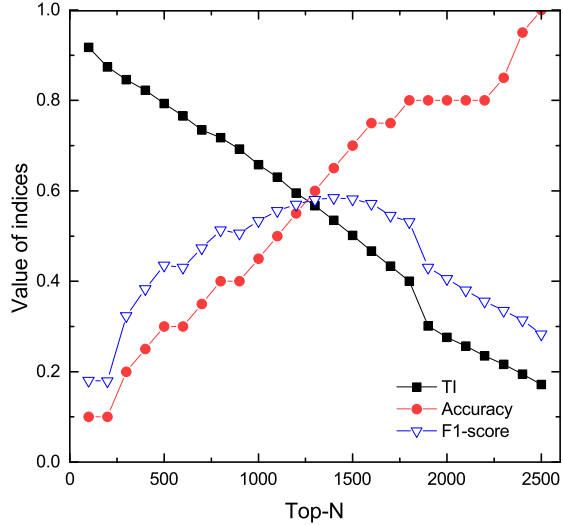}\\
  \caption{Time consumption compare on WMD and SimHash+WMD.}\label{Fig4}
\end{figure}

Figure \ref{Fig4} shows that $TI$ decreases with Top$-N$ increasing while the accuracy increasing with Top$-N$ increasing. Since the $F_1-score$ achieves maximum at the Top-1500, it is recommended to choose Top-1500 to screen potential  similar $Docs$ for 50\% time saving and 20\% precision reducing.
\

\begin{figure}
  \centering
  \includegraphics[width=10cm]{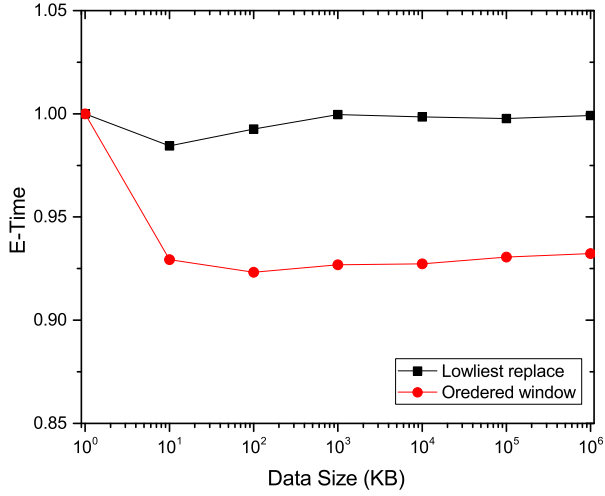}\\
  \caption{$E-Time$ on two different strategies in WMD+SimHash.}\label{Fig5}
\end{figure}

We also compare the time consumption of two improved screen strategies (\emph{lowliest replace elimination based strategy} and \emph{ordered window filling based strategy}) on DATA2. As shown in Fig. \ref{Fig5}, the \emph{lowliest replace elimination based strategy} only save 1\% time while the \emph{ordered window filling based strategy} can save about 7.5\% running time in screening process. The $E-Time$ index in Fig. \ref{Fig5} is a time index enhanced by the time consumption of all hamming distances sort strategy,
\begin{equation}\label{eq13}
   E-time=\frac{T}{T_0},
 \end{equation}
where $T$ is the time consumption of \emph{lowliest replace elimination based strategy} or \emph{ordered window filling based strategy}; $T_0$ is that of whole sort strategy.

Finally, we test the HINM-ESS's accuracy on relationship between entities using DATA1. We test on 8 $Topics$ to verify whether the $Items$ are linked to the right $Topics$ or not. The precision equals to the ratio of the number of test $Items$ and the number of correct $Items$. Generally, more training samples lead to higher precision. The precision on Topic $\sharp$4 and $\sharp$6 is higher than that on other Topics, since the number of training $Items$ in these two $Topics$ is about third times larger than that in others, as shown in Table \ref{table1}. In addition, the size of $Items$' training sample sets are larger than that of $Docs$'. Hence, as anticipated, the performance of precision on $Item-Topic$ reflecting is much higher than that of $Docs$.

\begin{table}
  \centering
  \caption{Accuracy of $Item-Topic$ relational model on DATA1.}\label{table1}
  \begin{tabular}{c cccc}
     \hline
     Topic & $\sharp$Training Item & $\sharp$Test Item & $\sharp$Correct Item & Precision(\%) \\ \hline
    $\sharp$1	&143	&17	&15	&88.24 \\
    $\sharp$2	&209	&25	&24	&96.00 \\
    $\sharp$3	&132	&16	&13	&81.25 \\
    $\sharp$4	&585	&67	&67	&100 \\
    $\sharp$5	&116	&14	&13	&92.86 \\
    $\sharp$6	&530	&60	&60	&100 \\
    $\sharp$7	&62	&8	&7	&87.50 \\
    $\sharp$8	&90  &11	&11	&100 \\  \hline
   \textbf{ Average}	&\textbf{233}	&\textbf{27}	&\textbf{26}	&\textbf{96.30} \\
     \hline
   \end{tabular}
\end{table}

\section{Conclusions}
Suffering from the complex and redundant equipment-standard system, a heterogeneous information network model for equipment-standard system(HINM-ESS) is presented in this paper to deal with some important issues in the system, such as standard documents searching, standard revising, standard controlling and, production designing. HINM-ESS contains three types of nodes that represent three types of entities and six types of entity relationships. $Doc-Doc$, $Doc-Item$, $Item-Item$, $Doc-Topic$, $Item-Topic$, and $Topic-Topic$ relational models are discussed and the detail modeling strategies are presented in this paper.

Experiments on two real data sets show that the modeling strategies are time saving and the accuracy is also rather good. Moreover, experiments show that HINM-ESS model can reflect real $Item-Topic$ relationships accurately when training sample is enough. Overall, HINM-ESS is an efficient and accurate model. It will provide a strong and firm support for applications in equipment-standard system.

We will consider the iterative strategy in the modeling process to improve the accuracy of relational models further in the future study.

\section*{Acknowledgements}
This work is partially supported by the National Natural Science Foundation of China under Grant Nos. 61433014 and 61673085, and by the Fundamental Research Funds for the Central Universities under Grant No. ZYGX2014Z002.
\section*{References}





\end{document}